\documentclass[a4paper]{JAC2003}
\usepackage{graphicx}
\usepackage{booktabs}
\usepackage{varwidth}
\usepackage{xcolor}

\begin{document}
\title{SIX-DIMENSIONAL WEAK--STRONG SIMULATIONS OF \\HEAD-ON BEAM--BEAM COMPENSATION IN RHIC\thanks{This work was supported by Brookhaven Science Associates, LLC under Contract No. DE-AC02-98CH10886 with the US Department of Energy.}}
\author{Y. Luo, W. Fischer, N.P. Abreu, X. Gu, A. Pikin,  G. Robert-Demolaize \\BNL, Upton, NY, USA }
\maketitle

\begin{abstract}
To compensate the large beam--beam tune spread and beam--beam resonance driving terms in the polarized proton operation in the Relativistic Heavy Ion Collider (RHIC), we  will introduce a low-energy DC electron  beam into each ring to collide head-on with the opposing proton beam. The device to provide the electron beam is called an electron lens. In this article, using a 6-D weak--strong beam--beam interaction simulation model, we investigate the effects of head-on beam--beam compensation with electron lenses on the proton beam dynamics in the RHIC 250~GeV polarized proton operation. This article is abridged  from the published article~\cite{RHIC-elens1}.
\end{abstract}

\section{Introduction}

To further increase the luminosity in the RHIC polarized proton (p-p) run, we plan to increase the proton bunch intensity with an upgraded polarized proton source~\cite{Anatoni}. In the 2012 RHIC 250~GeV p-p  run, the maximum bunch intensity at the beginning of physics store was $1.7\times10^{11}$. With the upgraded polarized proton source, we expect that the maximum bunch intensity will be increased up to $3.0\times10^{11}$.  Assuming the normalized r.m.s.\@ transverse emittance of 15~$\pi$ mm$\cdot$mrad, the linear incoherent tune shift or the beam--beam parameter will reach 0.03.

Currently, the working tune space for the RHIC p-p operation is chosen between 2/3 and 7/10 to achieve a good beam lifetime at store with beam--beam interaction and to maintain the proton polarization on the energy ramp and at the physics store~\cite{Mei1}. The 7/10 tune space is not only a 10th betatron resonance but also a spin depolarization resonance. Therefore, there is not enough tune space between 2/3 and 7/10 to hold the beam--beam generated tune spread when the bunch intensity is greater than  $2\times10^{11}$.

To reduce the beam--beam tune spread and also to compensate the non-linear beam--beam resonance driving terms, we plan to install head-on beam--beam compensation in the RHIC p-p operation. The proton beams collide at IP6 and IP8. A d.c. low-energy electron beam will be introduced into each ring around IP10 to head-on collide with the proton beam. The electron beam should have the same transverse profile as the proton beam.  The device to provide the electron beam for this purpose is called an electron lens (e-lens)~\cite{fermi-elens2}.

In the following,  with the 6-D weak--strong simulation, we study the head-on beam--beam compensation with e-lenses   on the proton beam dynamics in the RHIC 250~GeV p-p runs. The results from dynamic aperture and proton beam loss rate calculations are presented.  Key beam  parameters involved in this scheme are varied to search for the optimum compensation condition.  The sensitivity of head-on beam--beam compensation to  beam imperfections and beam offsets is also studied.

\begin{figure}
\center
\includegraphics[width=60mm]{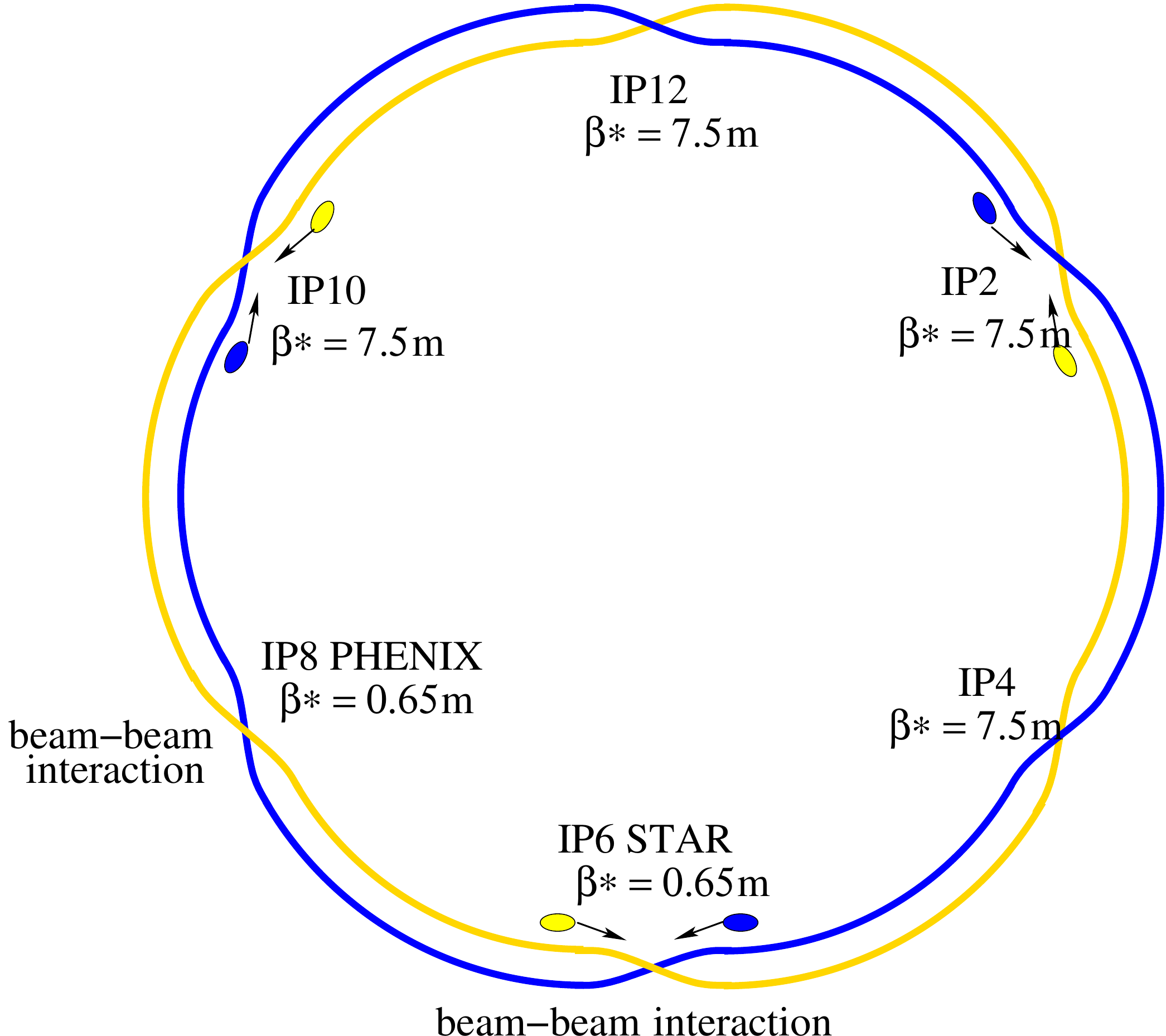}
\caption{The layout of the head-on beam--beam compensation in the RHIC. E-lenses are to be installed on either side of IP10.}
\end{figure}

\section{LATTICE AND BEAM PARAMETERS}

In the following simulation, we use a proposed Blue ring lattice for 250~GeV RHIC polarized proton operation. Table~\ref{tab1} lists the lattice and beam parameters.  The $\beta^*$ values at IP6 and IP8 are 0.5~m. The $\beta$ values at the e-lenses are 10~m.  The RHIC has not yet operated with $\beta^{*}=0.5$~m. In the 2012 RHIC 250~GeV polarized proton run,  we achieved $\beta^*=0.65$~m.  In this study, we assume that the r.m.s.\@ transverse emittance is 15~$\pi$ mm$\cdot$mrad and the
r.m.s.\@ relative momentum spread is $1.4\times10^{-4}$.

To cancel the non-linear beam--beam Resonance Driving Terms (RDTs) more effectively, the betatron phase advance between the beam--beam interaction and the e-lens should be $k\pi$, where $k$ is an integer. Since we only have one e-lens for each ring in the current design, we would like to have the betatron phase advances between IP8 and the e-lenses set at $k\pi$.  For the above lattice, the default phase advances between IP8 and the e-lens are (8.5$\pi$,  11.1$\pi$). In the following study, we will insert an artificial phase shifting matrix to bring them to (9$\pi$,  11$\pi$).

For simplicity, we  define that half and full beam--beam compensation compensate the half and full total linear incoherent beam--beam tune shift. Their compensation strengths are  0.5 and 1, respectively. If the electron beam has the same transverse r.m.s.\@ beam size as the proton beam at the e-lens,  we have  $N^*_\mathrm{e} = N_\mathrm{p}$ and $N^*_\mathrm{e} = 2N_\mathrm{p}$ for  half and full  beam--beam compensation, respectively. Here, $N^*_\mathrm{e}$  and $N_\mathrm{p}$ are the electron populations in the e-lens and the proton bunch intensity.

\begin{table}
\centering
\caption{The lattice and beam parameters used in this study}
\begin{tabular}{lcc}
\hline \hline
Parameter                                   &  Value  \\  \hline
Circumference                          & 3833.8451~m \\  
Energy                                      & 250~GeV                \\  
Working point                          &  $(28.67,29.68)$   \\  
Linear chromaticities                    &  $(1,1)$ \\ 
Second-order chromaticities              &  (2800, 2900) \\  
Transverse r.m.s.\@ emittance      & 2.5~mm$\cdot$mrad \\  
$\beta^*_{x,y}$ at IP6 and IP8              & 0.5~m  \\  
$\beta_{x,y, e-lens}$ at the e-lens               & 10~m   \\  
Trans. r.m.s.\@ beam size at IP6 and IP8                   & 68~$\mu$m   \\  
Trans. r.m.s.\@ beam size at e-lens                        & 310~$\mu$m \\  
$\Delta \Phi_{x,y}$ between IP6 and IP8      & (10.6$\pi$, 9.7$\pi$)\\  
$\Delta \Phi_{x,y}$ between IP8 and the e-lens   & (8.5$\pi$,  11.1$\pi$)\\  
RF harmonic number                          & 360                  \\  
RF cavity voltage                       & 300~kV               \\  
Longitudinal r.m.s.\@ bunch area              &  0.17~eV$\cdot$s              \\  
Bucket height            &   $1.1\times10^{-3}$  \\  
Relative r.m.s.\@ momentum spread               &  $1.4\times10^{-4}$  \\  
R.m.s.\@ bunch length                        &  0.45~m             \\
\hline \hline
\end{tabular}
\label{tab1}
\end{table}

\section{The Simulation Model}

In the following simulation study, we track the proton particles element by element~\cite{SimTrack}. The non-linear magnetic field errors in the triplets and separation dipoles in the interaction regions are included. Each magnetic element is modelled with a 6-D symplectic transfer map. We have adopted
fourth-order symplectic integration~\cite{4thSI}. To save time in the long-term particle tracking, we model the magnetic multipoles as thin lens kicks. Tunes and chromaticities are rematched before tracking.

Considering that $\beta^{*}$ is comparable to the proton bunch length,  we use the 6-D weak--strong synchro-beam map~\cite{6DBB} to model the proton-proton  beam--beam interaction  at IP6 and IP8.  The strong bunch is split into 11 slices to achieve good convergence.  Considering that the e-lens is working in a d.c.\@ mode, its electric and magnetic fields are static. In the simulation code, we split the  2~m long e-lens into eight slices. Each slice is modelled as a drift -- a 4-D  weak--strong  beam--beam kick.  The 4-D  weak--strong  beam--beam kick is given by Bassetti and Erskine~\cite{4DBB}.

To fully  use the available tune space between 2/3 and 7/10 and for better comparison of the simulation results under different beam--beam conditions,  we fix the zero-amplitude particle tunes at (0.67, 0.68) under different beam--beam conditions,   except in the proton working point scan. The RHIC polarized proton operational experience shows that a lower working point between 2/3 and 7/10 is preferable to obtain a better beam--beam lifetime and to  preserve the proton polarization at store.  In the simulation, the linear chromaticities are set to (1,1).

\section{CALCULATION OF THE DYNAMIC APERTURE}

In this section, we calculate the proton dynamic aperture with head-on beam--beam compensation in the RHIC.  Particles are tracked in 10 phase angles in the ($x,y$) plane up to $10^{6}$ turns. The initial relative momentum error is $0.42\times10^{-4}$. We compare the minimum  dynamic aperture under different beam and lattice conditions. The dynamic aperture is given in units of r.m.s.\@ transverse beam size $\sigma$.

Figure~2 shows the dynamic apertures without,  with half, and with full head-on beam--beam compensation.  The proton bunch intensity varies from $1.0\times10^{11}$ to $3.0\times10^{11}$.  In this calculation, the betatron phase advances between IP8 and the e-lens are the default ones (8.5$\pi$,  11.1$\pi$). From Figure~2, half beam--beam compensation increases the dynamic aperture when the proton bunch intensity is bigger than $2.0\times10^{11}$. Full beam--beam compensation reduces the  dynamic aperture for all shown bunch intensities.

Figure~3 shows the dynamic apertures versus the head-on beam compensation strength. In this study, we keep  the electron transverse beam size the same as the proton beam size at the e-lens, and adjust the electron beam intensity to change the beam--beam compensation strength.  From Figure~3, the proton dynamic apertures drop sharply when the compensation strength is larger than 0.7.  The optimized compensation strengths for the bunch intensities $2.5\times10^{11}$ and $3.0\times10^{11}$ are around 0.5--0.6.

Figure~4 shows the dynamic apertures of half head-on beam--beam compensation with $k\pi$ phase advances between IP8 and the e-lens and the second-order chromaticity correction. The second-order chromaticities without correction are around 2800.  With correction, they are below 500.  The results show that the  $k\pi$ phase advances and second-order chromaticity further improve the dynamic aperture of  half beam--beam compensation by about 1~$\sigma$ for all the bunch intensities shown in the plot. In the above calculation,  the zero-amplitude tunes of the proton beam are fixed at (0.67, 0.68).  With beam--beam compensation, the tune footprint becomes smaller and it is possible to scan  the proton working point between 2/3 and 7/10 to maximize the dynamic aperture with a  better working point.  Figures~5 and ~6  show the dynamic apertures of half and full beam--beam compensation in the tune scan. The horizontal axis is the fractional horizontal zero-amplitude  tune. The fractional vertical zero-amplitude tune is always 0.01 higher than the horizontal one.  Simulation results show that half beam--beam compensation prefers a lower working point, between 2/3 and 7/10,  while full beam--beam compensation prefers a higher working point.  The maximum dynamic aperture of half beam--beam compensation in the tune scan is higher than that with full beam--beam compensation.

\begin{figure}
\centering
\includegraphics[width=75mm]{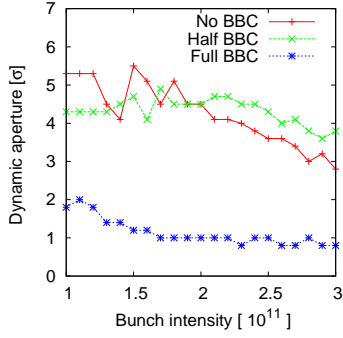}
\caption{Dynamic apertures without beam--beam compensation and with half and full beam--beam compensation. }
\label{fig:da_1}
\end{figure}

\begin{figure}
\centering
\includegraphics[width=75mm]{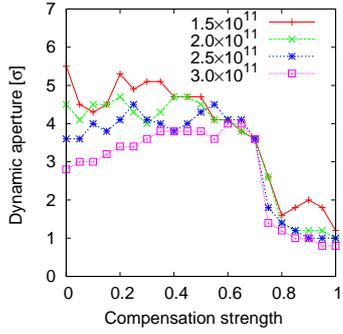}
\caption{Dynamic apertures of four proton bunch intensities versus the head-on beam--beam compensation strength. }
\label{fig:da_2}
\end{figure}

\begin{figure}
\centering
\includegraphics[width=75mm]{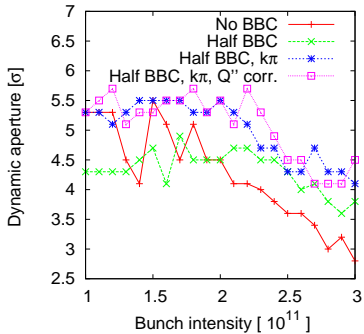}
\caption{Dynamic apertures of half head-on beam--beam compensation with the betatron phase advance adjustment  and the global second-order chromaticity correction. }
\label{fig:da_3}
\end{figure}

\begin{figure} [!]
\centering
\includegraphics[width=75mm]{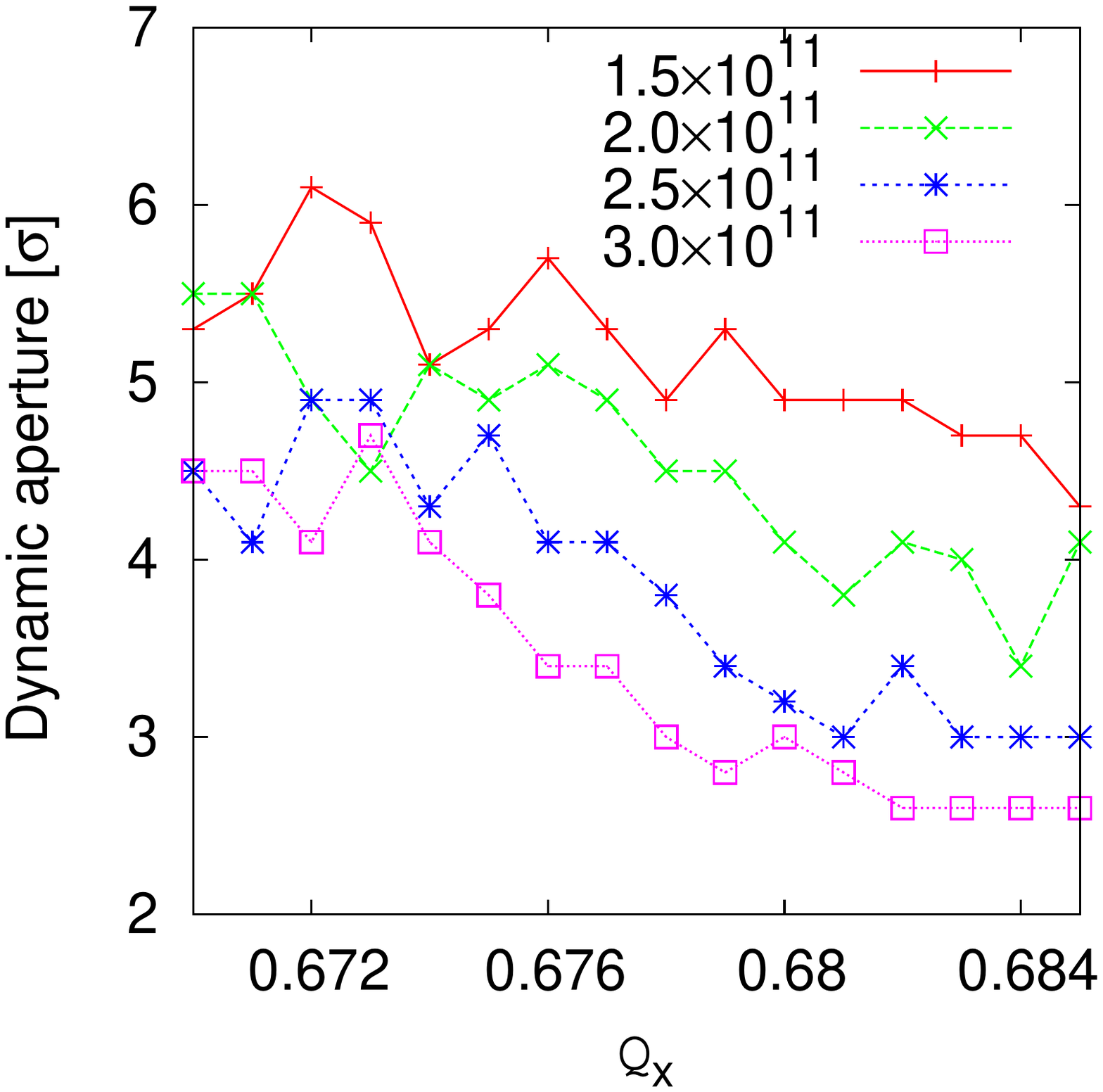}
\caption{Dynamic apertures of half head-on beam--beam compensation in the scan of the proton working point. }
\label{fig:da_5}
\end{figure}

\begin{figure}  [!]
\centering
\includegraphics[width=75mm]{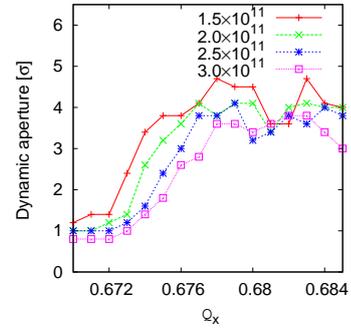}
\caption{Dynamic apertures of full head-on beam--beam compensation in the scan of the proton working point. }
\label{fig:da_6}
\end{figure}

\section{CALCULATION OF THE PARTICLE LOSS RATE}

In this section, we calculate the proton beam loss rate with multi-particle tracking in the presence of head-on beam--beam compensation. Limited by computing capacity, in  most cases we track 4800 macro-particles  up to  $2\times10^6$  turns. $2\times10^6$  turns  are 24 s for the RHIC.

Particles with large transverse amplitudes and large momentum deviations are probably lost in long-term tracking. However, for a limited number of macro-particles sampled from a solid Gaussian distribution, there are only a few macro-particles in the Gaussian bunch tail.  To detect a small beam loss in  $2\times10^6$  turns without increasing the number of macro-particles, we  track particles  with an initially hollow Gaussian distribution.

In this approach, we assume that the particles in the bunch core are stable and will not be  lost in $2\times10^6$ turns. To save computing time, we only track macro-particles the transverse or longitudinal amplitudes of which are bigger than a certain r.m.s.\@ beam size. The boundary between the stable core and the unstable bunch tail  is carefully chosen. We first calculate the dynamic aperture and set the boundary well below it.

Figures~7--9 show the relative proton beam losses in  $2\times10^{6}$ turns under different beam--beam compensation conditions with proton bunch intensities $2.0\times 10^{11}$, $2.5\times 10^{11}$, and $3.0\times 10^{11}$.   Just as in the dynamic aperture calculation, here we set the  zero-amplitude tunes of the proton beam to (0.67, 0.68) and the linear chromaticities to (1,1). 

For each proton bunch intensity, we compare the relative proton beam losses without beam--beam compensation, with half beam--beam compensation, with the optimized betatron phase advances $k\pi$ between IP8 and the e-lens,  and with the global second-order chromaticity correction.  For all the  three bunch intensities, full head-on beam--beam compensation gives a much bigger beam loss than other beam--beam conditions and therefore its beam loss is not plotted.

From Figs.~7--9, half head-on beam--beam compensation reduces proton particle losses  with bunch intensities $2.5\times 10^{11}$ and $3.0\times 10^{11}$ in $2\times10^{6}$ turns. Also, the $k\pi$ phase advances between IP8 and the e-lens and the second-order chromaticity correction further improve the proton beam lifetime, which agrees the results from above dynamic aperture calculations. For the bunch intensity  $2.0\times 10^{11}$, simulation shows that head-on beam--beam compensation does not increase the proton lifetime.

\begin{figure} [!]
\centering
\includegraphics[width=75mm]{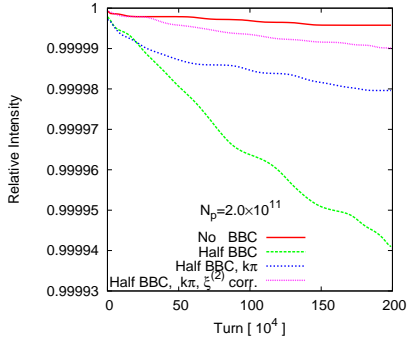}
\caption{Relative proton beam losses in $2\times10^6$ turns for the proton bunch intensity $2.0\times10^{11}$.}
\label{lifetime_1}
\end{figure}

\begin{figure} [!]
\centering
\includegraphics[width=75mm]{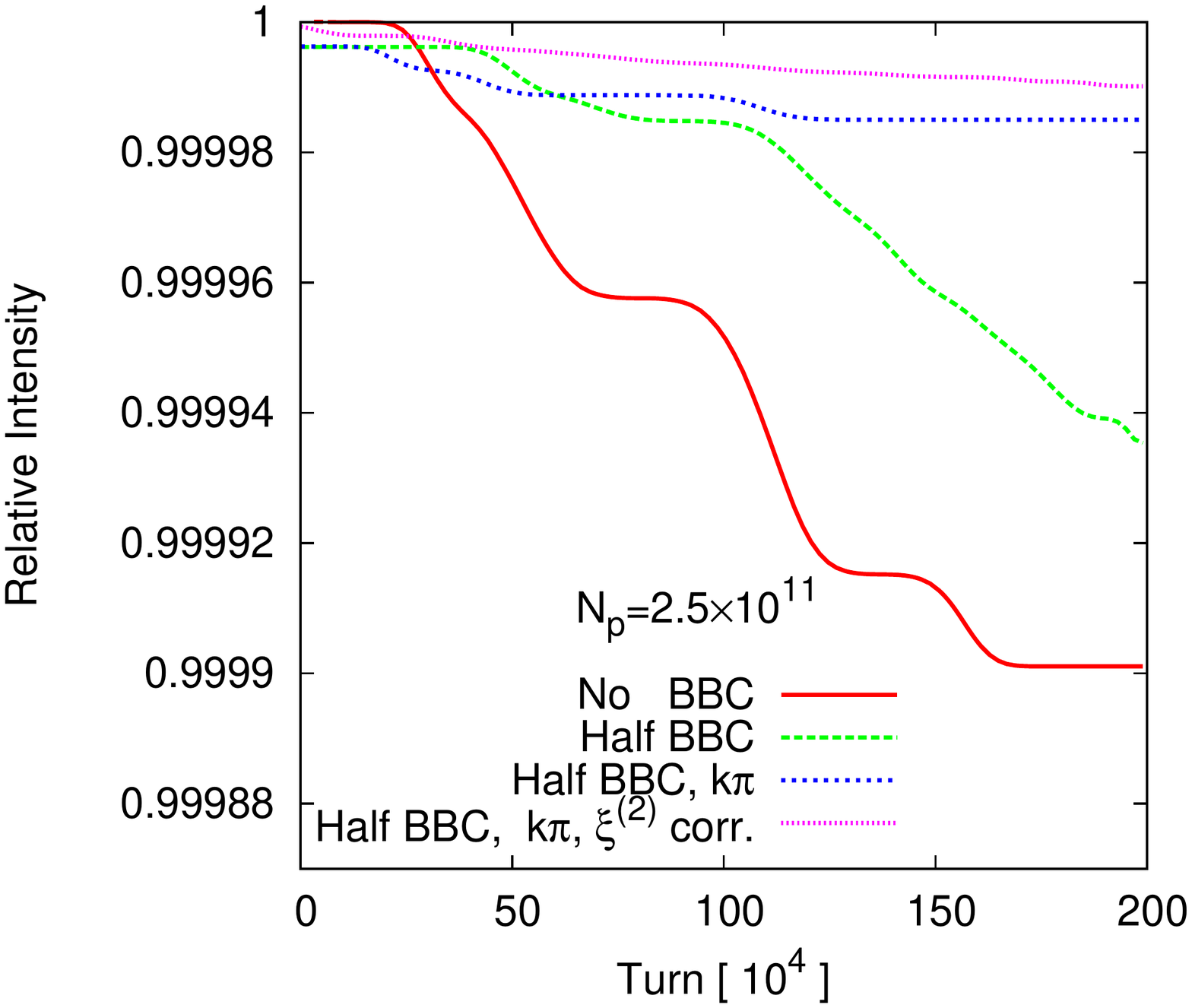}
\caption{Relative proton beam losses in $2\times10^6$ turns for the proton bunch intensity $2.5\times10^{11}$.}
\label{lifetime_2}
\end{figure}

\begin{figure} [!]
\centering
\includegraphics[width=75mm]{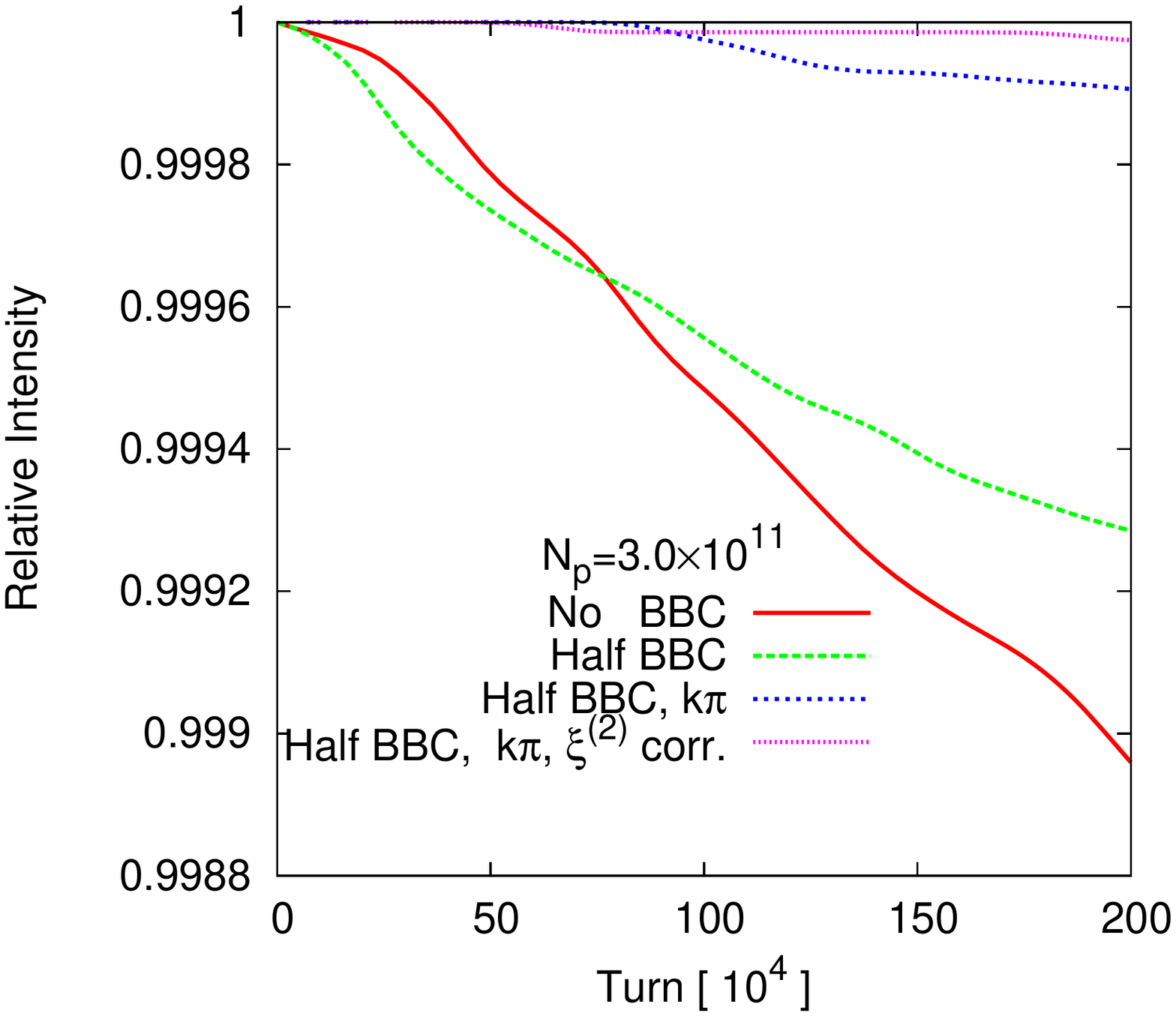}
\caption{Relative proton beam losses in $2\times10^6$ turns for the proton bunch intensity $3.0\times10^{11}$.}
\label{lifetime_3}
\end{figure}

\section{Sensitivity Study}

\begin{figure}[!]
\centering
\includegraphics[width=75mm]{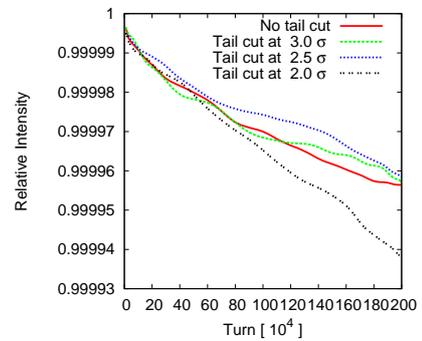}
\caption{The relative proton beam loss with the truncated Gaussian tail of the electron beam. }
\label{tolerance4}
\end{figure}

\begin{figure} [!]
\centering
\includegraphics[width=75mm]{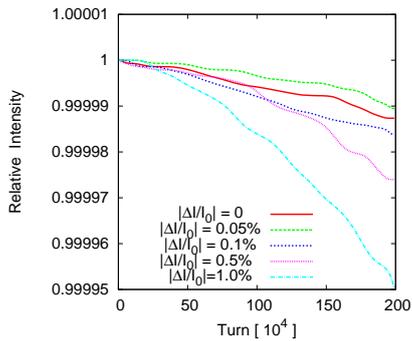}
\caption{The relative proton beam loss versus the random noise in the electron beam current. }
\label{tolerance3}
\end{figure}

\begin{figure} [!]
\centering
\includegraphics[width=75mm]{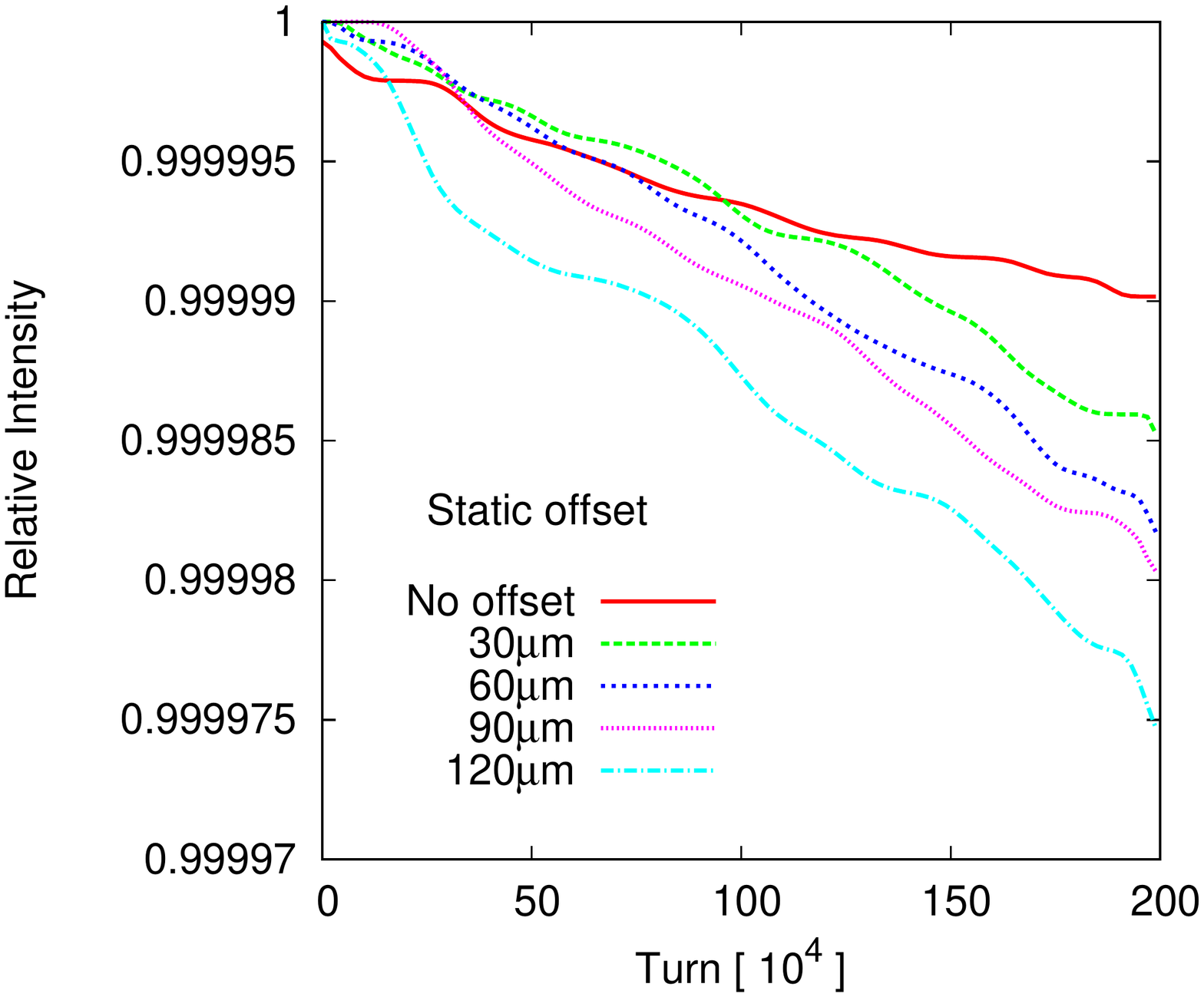}
\caption{The relative proton beam loss with static transverse offset between the e-lens and the proton beam in the e-lens.}
\label{tolerance6}
\end{figure}

\begin{figure} [!]
\centering
\includegraphics[width=75mm]{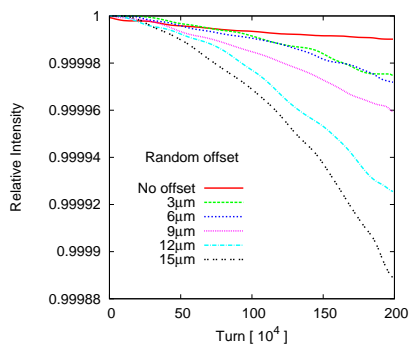}
\caption{The relative proton beam loss with random transverse offset between the e-lens and the proton beam in the e-lens.}
\label{tolerance7}
\end{figure}

In this section, we study the sensitivity of head-on beam--beam compensation to the beam imperfections and beam noise. We focus on the  Gaussian tail truncated electron beam, the random noise in the electron beam current, and the static and random offsets between the electron and  proton beams. The baseline for this study is with the proton bunch intensity $N_\mathrm{p}=2.5 \times 10^{11}$ and half beam--beam compensation. The betatron phase advance adjustment and second-order chromaticity correction are included.

In the above simulation study, we assume that the electron beam has a perfect transverse Gaussian distribution.  Simulation of the electron gun for the RHIC e-lens system shows that the electron beam has a Gaussian tail cut off at 2.8~$\sigma$. Figure~10 shows the calculated relative proton beam loss with the electron beam's  Gaussian tail cut off at 3~$\sigma$, 2.5~$\sigma$ and 2~$\sigma$. Compared to the baseline with a perfect Gaussian distribution, the Gaussian tail cut at  2.8~$\sigma$ from the current electron  gun design is acceptable.

Due to the instability of the power supplies of the electron gun, there is noise in the electron beam current.   Figure~11 shows the relative proton beam loss versus the random electron current noise.  The proton beam loss with a random noise below  0.1\%  in  the electron current is comparable to the baseline  without current noise.  In the design of the RHIC electron gun system, we require that the stability of the power supplies of the electron gun should be better than 0.1\%.

Overlapping of the electron and proton beams in the e-lens plays a crucial role in head-on beam--beam compensation. Figures~12 and 13 show the calculated relative proton beam losses with static and random offsets between the electron and proton beams.  Based on the simulation results, in the RHIC e-lens design, we set the tolerance of the static offset error to 30~$\mu$m, which is a 10th of a r.m.s.\@ beam size in the e-lens, and the random offset to  9~$\mu$m, which requires the stability of the bending magnet's power supply to be better than 0.01\%.

\section{SUMMARY}

In this article, with a 6-D weak--strong beam--beam model, we have investigated the effects of head-on beam--beam compensation with e-lenses on the proton beam dynamics in the RHIC 250~GeV p-p operation.  We found that half beam--beam compensation improves the proton dynamic aperture and beam lifetime.  The $k\pi$ phase advances between IP8 and the e-lens, and the global second-order chromaticity, further increase the proton dynamic aperture and particle loss rate.  The sensitivity of half beam--beam compensation on the electron profile, the electron current, and the overlapping of the electron and proton beams are studied and their tolerances are set.


\end{document}